\documentclass[runningheads]{llncs}
\usepackage[T1]{fontenc}
\usepackage{graphicx}
\usepackage{algorithm,algcompatible}
\usepackage{physics}
\usepackage[english]{babel}
\usepackage[square,numbers]{natbib}
\usepackage{amsfonts}
\usepackage{caption}
\usepackage{subcaption}
\usepackage{inconsolata}

\begin{document}
\title{Multi-Channel Currency: A Secure Method Using Semi-Quantum Tokens}
\titlerunning{Multi-Channel Currency}
%
\author{Yichi Zhang\inst{1}\orcidID{0000-0003-4658-6388} \and
Siyuan Jin\inst{1}\orcidID{0000-0001-7971-0424} \and
Yuhan Huang\inst{1}\orcidID{0000-0002-7950-3895} \and
Qiming Shao\inst{1}\orcidID{0000-0003-2613-3031}}
\authorrunning{Y. Zhang et al.}
%
\institute{The Hong Kong University of Science and Technology, Clear Water Bay, Kowloon, Hong Kong
\email{\{lionel.zhang, siyuan.jin, yhuangfv\}@connect.ust.hk, eeqshao@ust.hk}}
\maketitle              
\begin{abstract}
Digital currencies primarily operate online, but there is growing interest in enabling offline transactions to improve digital inclusion. Existing offline methods struggle with double-spending risks, often limiting transaction amounts. In this work, we propose a quantum-state-based currency system that uses the non-cloning theorem to enable secure, multi-channel transactions without the risk of double spending. We demonstrate this system's implementation with experimental results, including use cases for currency transfers and swaps. To mitigate credit risks in swaps, we also integrate blockchain to show its wide applicability. Our approach paves the way for quantum-secure digital currencies and opens new possibilities for optimizing multi-channel tokens.

\keywords{Quantum Money  \and Blockchain \and DeFi.}
\end{abstract}

\section{Introduction}

The rapid digitalization of financial systems has prompted significant interest in payment innovations that enhance accessibility and security, particularly in regions with limited infrastructure~\cite{ozili2021financial}. Yet, existing IS research on digital tokens and decentralized payment systems has not adequately addressed offline and multi-channel transaction scenarios, which are critical for financial inclusion.

Traditional payment systems, built on a single-ledger model to prevent double-spending, rely heavily on online synchronization. This reliance becomes problematic in areas with poor internet connectivity, leading to inefficiencies such as limited transaction capacity, and increased fraud risks, rendering these systems unsuitable for offline environments~\cite{karame2015misbehavior}. To address these issues, researchers have proposed solutions such as limiting transaction amounts or using micropayments, implementing specialized payment designs~\cite{daza2015frodo}, and deploying double-spending attack detection mechanisms~\cite{tso2017off}.

Yet, existing measures fall short of fully addressing the persistent threat of double-spending attacks in digital token systems. This paper introduces a groundbreaking token protocol that enables secure multi-channel transactions without requiring real-time synchronization. Digital tokens, which represent ownership of digital or physical assets, have revolutionized how ownership information is conveyed, reducing costs and complexity~\cite{ziolkowski2020decision}. Building on this foundation, semi-quantum tokens leverage quantum technology to achieve unprecedented security features, including unclonability and secure destruction, which are unattainable with classical tokens. By eliminating dependence on continuous online connectivity, the proposed system significantly expands accessibility to regions with limited infrastructure. Crucially, this approach establishes a robust solution to prevent double-spending in offline and decentralized environments, addressing a fundamental challenge in digital payment systems. 

We conduct experiments to validate the proposed protocol. The protocol demonstrates strong resistance to double-spending, even in offline environments. We also demonstrate how it can work combining with public blockchain and smart contracts.

This work makes the following contributions. First, we propose that quantum money can be an effective method to ensure multi-channel minting and transactions, ensuring robustness against counterfeiting. Second, we show how quantum money effectively prevents double-spending in offline transactions, providing a secure foundation for decentralized payment systems. Together, these features enable a more reliable and inclusive financial ecosystem, particularly in underserved regions.

The proposed protocol has significant practical implications. It offers a secure and scalable solution for offline payment systems, reducing fraud and enhancing trust. By enabling reliable financial transactions in areas with poor connectivity, the system fosters financial inclusion and economic participation in underserved communities. Ultimately, this work contributes to the development of resilient and accessible payment infrastructures, bridging the gap between technological advancements and practical financial needs.

\section{Preliminary}

\subsection{Quantum Token}

By combining classical homomorphic encryption (HE) with quantum encryption techniques, quantum fully HE provides a secure framework for outsourcing quantum computations and enable local minting. Its hybrid nature ensures compatibility with classical systems while enabling robust quantum privacy protections. The detailed techniques is documented in Appendix \ref{crptography tools}.

\subsection{Quantum Lighting and Non-Cloning Theorem}

The work of \cite{shmueli2022public} proves that the minting circuit for quantum tokens satisfies the property of \textit{quantum lightning}, as defined in \cite{zhandry2021quantum}. Quantum lightning is a cryptographic concept that ensures the uniqueness and non-clonability of quantum states produced by a specific quantum process. This property is foundational for protocols that rely on quantum tokens, as it ensures that no two identical quantum tokens can exist, thereby preventing counterfeiting. The detailed techniques is documented in Appendix \ref{crptography tools}.

The \textit{non-cloning theorem} is a fundamental result in quantum mechanics that provides a natural form of security for quantum information. It states that it is impossible to create an exact duplicate (or clone) of an arbitrary, unknown quantum state. This theorem was first formalized in \cite{wootters1982single, dieks1982communication} and is a cornerstone of quantum cryptography.


This principle has profound implications for cryptographic protocols involving quantum tokens:
\begin{itemize}
    \item \textbf{Uniqueness of Quantum Tokens:} The non-cloning theorem ensures that quantum tokens cannot be copied or counterfeited, as duplicating the quantum state associated with a token would violate the theorem.
    \item \textbf{Physical-Level Security:} Unlike classical digital tokens, which can be duplicated unless additional cryptographic protections are in place, quantum tokens derive their security directly from the laws of quantum mechanics. This makes them inherently resistant to forgery.
\end{itemize}

In essence, the non-cloning theorem provides a foundational layer of security for quantum cryptographic systems, ensuring that any attempt to duplicate or tamper with a quantum token will fail. This property is crucial for maintaining the integrity of protocols that rely on the uniqueness and authenticity of quantum states, such as semi-quantum payments or quantum lightning-based systems.

\subsection{Problem Statement}

A double-spending attack is a major security threat in digital token systems~\cite{osipkov2007combating}, where the same token is fraudulently used in multiple transactions. Unlike physical currencies, digital tokens can be copied or manipulated without proper safeguards. Traditional systems rely on centralized ledgers or blockchain technology to prevent double-spending by synchronizing transactions across the network~\cite{jin2022cev}. However, in offline or multi-channel environments, where real-time synchronization is unavailable, ensuring token uniqueness becomes significantly more challenging. In such settings, users may attempt to spend the same token across independent channels simultaneously, exploiting the lack of coordination between channels. Addressing this issue is critical to maintaining the security and reliability of digital payment systems, especially in decentralized or offline scenarios.

\section{Methodology}
\begin{figure}[h!]
    \centering
    \includegraphics[width=1\linewidth]{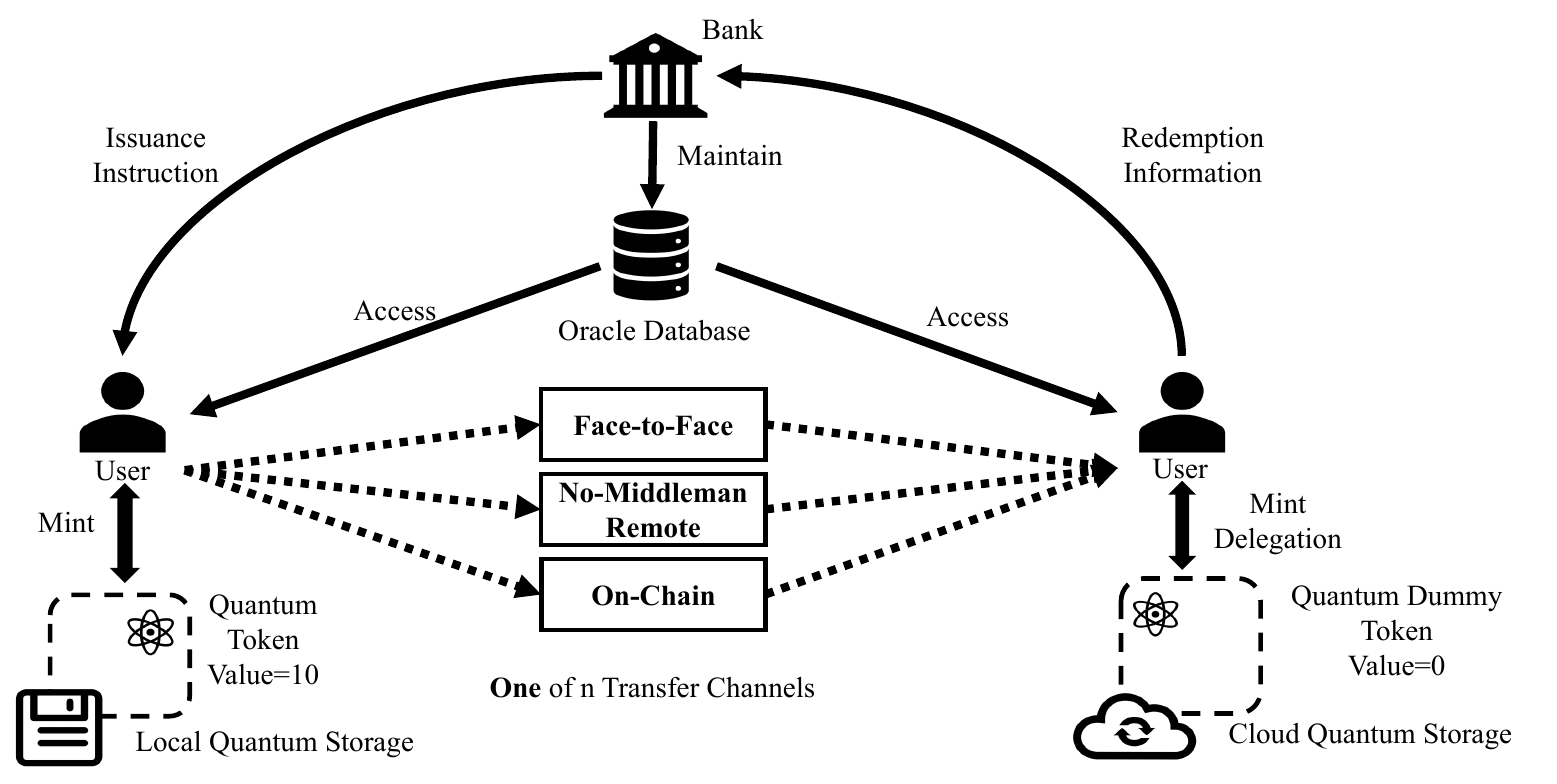}
    \caption{Multi-Channel Quantum Currency Ecosystem}
    
    \label{fig:ecosystem}
\end{figure}

The proposed multi-channel semi-quantum currency protocol is an IT artifact designed to mitigate double-spending in offline payment environments. The overall system is demonstrated in Figure~\ref{fig:ecosystem}. We then explain each operations in the following subsections.

\subsection{Semi-Quantum Token}\label{semi-quantum token}

We use \cite{shmueli2022semi} as the semi-quantum token entity of our system. According to the protocol, a pure classical bank without any quantum resources can delegate a mint mission to a quantum receiver or a classical receiver who can access a 3rd-party quantum cloud \cite{zhang2024cloud} through the classical channel. After mint, the receiver will attain an unclonable quantum state which servers as a quantum token, while the bank will public a set of oracles that ensures everyone who can access the quantum token can verify the unit's authenticity. The quantum token holder can also transfer the quantum token's value to another one through the classical channel by certificated destruction.

\textit{Certificated destruction is the nature of semi-quantum token.} When transferring the value from token $A$ to dummy token $B$, the holder of $B$ first sends the unique index $id_B$, which is assigned from the bank when each token is minted, to the holder of token $A$. Then the holder of token $A$ will sign the value of token $A$ to dummy token $B$ by certificated destruction. The certificated destruction method means the set of quantum token units is measured according to the bits in $id_B$ respectively and cannot be reversed into the original quantum states. We call the classical measurement result the signature $\sigma_{B\leftarrow A}$. The signature is the unforgeable proof that the value of token $A$ has been transferred into the dummy token $B$ and anyone or algorithm can verify the correctness of the transaction via the public oracle database. Meanwhile, the former holder of token $A$ cannot double-spend the token twice because the token has been destroyed before and cannot sign to another token. In addition, the holder of another token $C$ is not able to claim the transaction because the index $id_C$ is different from $id_B$.

\subsubsection{Minting a Quantum Token}

\begin{figure}[h]
    \centering
    \includegraphics[width=1\linewidth]{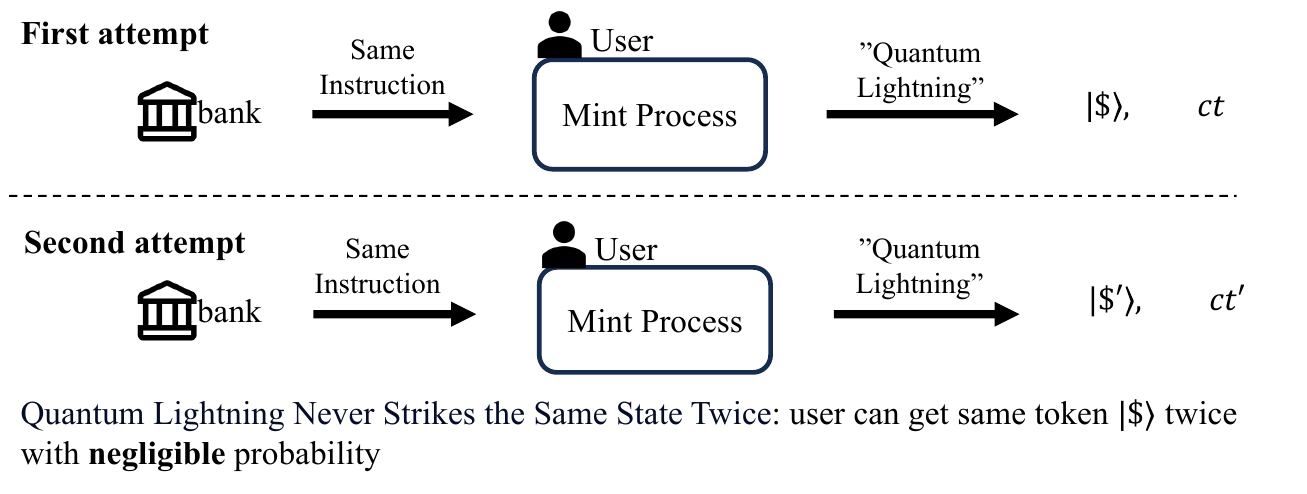}
    \caption{Quantum Money Minting Process}
    \label{fig:quantum_money_mint}
\end{figure}

The minting process is the first step in creating a quantum token. This process ensures that the token is unique and unclonable, leveraging the quantum property of Quantum Lightning, which prevents duplication of certain quantum states. Figure~\ref{fig:quantum_money_mint} illustrates the quantum money minting process, highlighting the core property of Quantum Lightning, which ensures the uniqueness of quantum tokens. In the first attempt, the user interacts with the minting process, which outputs a quantum state ($\ket{\$}$) and a classical tag ($ct$). If the user repeats the same instruction, the process generates a different quantum state ($\ket{\$^{`}}$) and tag ($ct^{`}$). This property ensures that the same quantum state cannot be reproduced with more than a negligible probability, preventing duplication. The bank assigns value only to the first reported token, ensuring the security and integrity of the system. This process demonstrates the unforgeability of quantum states, a critical feature for quantum money systems.

During minting, the user interacts with the system and receives a quantum state denoted as $\ket{\$}$ and a corresponding classical tag $ct$. If the same minting instructions are repeated, the system will produce a completely different quantum state $\ket{\$'}$ and tag $ct'$. This behavior ensures that no two identical tokens can be created, making it impossible to duplicate tokens. The bank assigns value to the first reported quantum token and ensures that any later attempts to report duplicate tokens are rejected. This mechanism provides strong security guarantees and ensures the uniqueness of each quantum token.

A quantum token is composed of two main parts:
\begin{itemize}
    \item \textbf{Classical Information:} Includes:
    \begin{itemize}
        \item Public keys ($pk_1, pk_2, ..., pk_n$), where each $pk_i$ corresponds to a quantum token unit.
        \item A value ($x$) assigned by the bank.
        \item A unique identifier ($id$), represented as a $\lambda$-length bit string to uniquely identify each token.
    \end{itemize}
    \item \textbf{Quantum Information:} A collection of quantum states ($\ket{\$}_1, \ket{\$}_2, ..., \ket{\$}_n$), where each quantum state corresponds to a unit of the token. These quantum states are unclonable due to the fundamental properties of quantum mechanics.
\end{itemize}

This minting process ensures that the quantum token is securely tied to the classical information publicly recorded by the bank, making it both unclonable and verifiable.

\subsubsection{Verification of a Quantum Token}

Quantum tokens can be verified without being destroyed, which allows the token holder to prove authenticity multiple times without losing the token itself. The verification process relies on the oracles published by the bank during minting. These oracles are publicly accessible and allow anyone to check whether a quantum token is valid. If a quantum token unit (e.g., $\ket{\$}$) passes verification, the quantum state remains intact and can continue to be used or transferred. 

The verification process can be summarized as follows:
\begin{itemize}
    \item The receiver tests the quantum token against the oracles published by the bank to check its validity.
    \item If the token passes these checks, it is accepted as authentic.
    \item If it fails, the token is rejected.
\end{itemize}

This ability to verify tokens without destroying them ensures that quantum tokens can be reused and remain practical for real-world applications.

\subsubsection{Transferring a Quantum Token}\label{sec:transfer}
The transfer process involves certified destruction, which ensures that the original token is invalidated once its value is transferred to the recipient's token. This mechanism prevents double-spending or unauthorized duplication.

The transfer process is based on a fundamental property of quantum mechanics: a quantum token unit can "collapse" into one of two distinct subspaces, depending on a classical control bit $b$. Once the token collapses, it cannot be restored to its original quantum state. This collapse serves as proof of destruction during the transfer process. The transfer works as follows:
\begin{enumerate}
    \item The sender (e.g., Alice) wishes to transfer the value of their quantum token $A$ to the receiver (e.g., Bob), who holds a dummy token $B$ with no value assigned yet.
    \item Alice signs each quantum token unit in $A$ according to Bob's token identifier ($id_B$) and produces a signature, denoted as $\sigma_{B \leftarrow A}$.
    \item This signature proves two things:
    \begin{itemize}
        \item The original token $A$ has been destroyed.
        \item The value of $A$ has been transferred to $B$.
    \end{itemize}
    \item Bob can then use the signature to prove that the transfer was successful and that the value of $A$ is now associated with $B$.
\end{enumerate}

The certified destruction of the original token ensures that each quantum token unit can only be destroyed once, and the transfer is irreversible and verifiable. This property makes quantum tokens highly secure for transferring value or ownership between parties.

\subsection{Multi-Channel Transactions}
\begin{figure}[h!]
    \centering
    \includegraphics[width=1\linewidth]{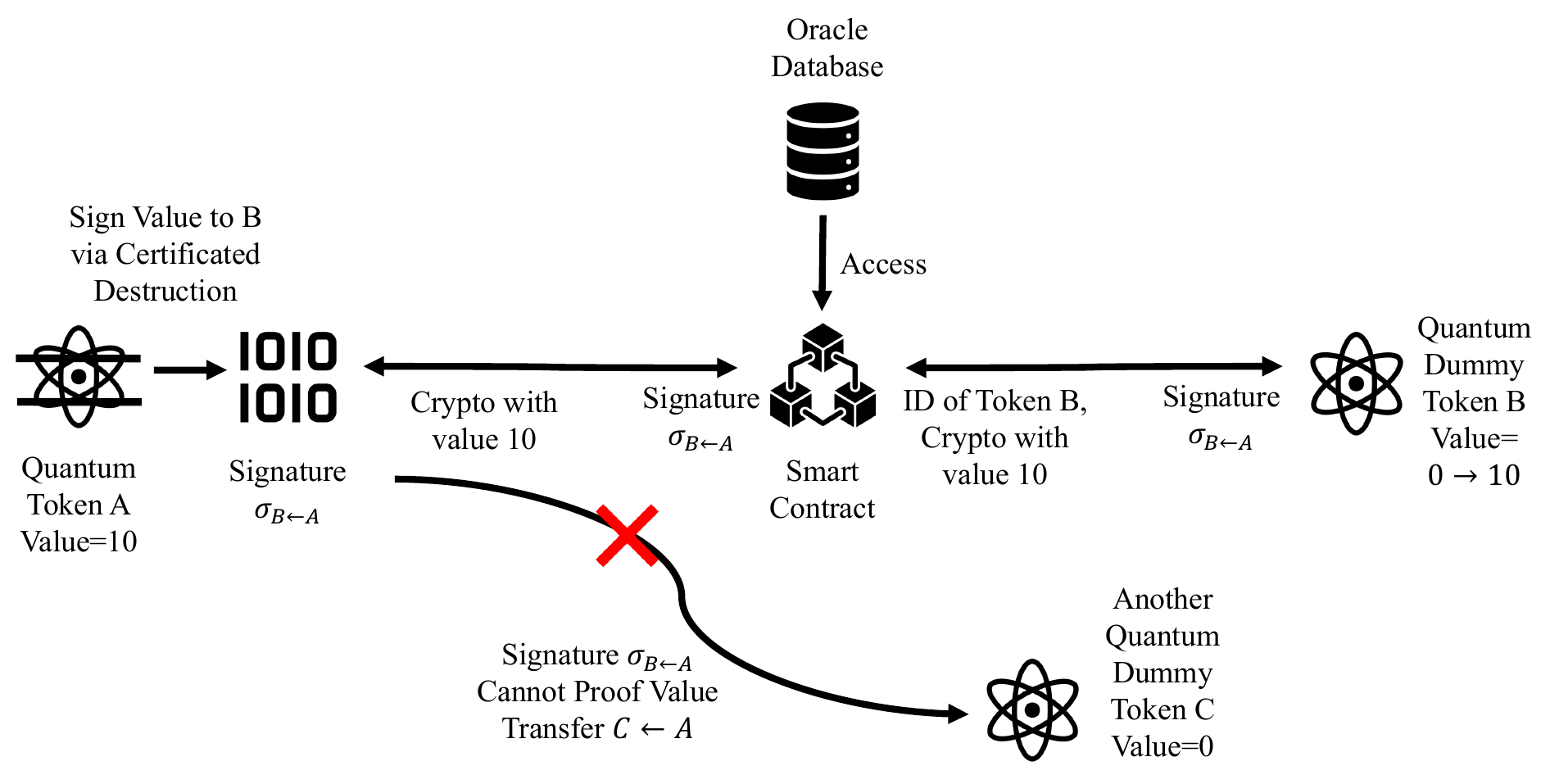}
    \caption{Signature via Certificated Destruction}
    
    \label{fig:anti_ds}
\end{figure}

Our multi-channel semi-quantum currency protocol supports face-to-face, no-middleman remote, and on-chain transfer. Since each channel has different advantages and disadvantages, the participants of the deal can choose any appropriate channel in different scenarios without taking the risk of double-spending. Fig. \ref{fig:anti_ds} is a typical diagram shows the dataflow of on-chain transfer method and how certificated destruction prevent double-spending.

\textbf{Face-to-Face Transfer} is the most trivial channel. The sender exchanges value of quantum token for goods via standard certificated destruction and verification. The procedure involves no third party nor central ledger. Meanwhile, the channel requires the two participants to meet locally which is inconvenience.

\textbf{No-Middleman Remote Transfer:}
Alternatively, the receiver can send the index of dummy token $B$ to the sender through classical channel and the sender sends back the signature $\sigma_{B\leftarrow A}$ through classical channel too. In this transfer channel, the sender and the receiver need not to meet locally but can also conduct the transfer without the participative of 3rd-party middleman (e.g. central bank, distributed ledger). The disadvantage of this channel is, after transfer, there is no mechanism ensures the receiver to return the goods with the corresponding value.

\textbf{On-chain Transfer:} In order to make up for the disadvantage of the previous channel, we propose the on-chain channel. We use smart contract to automatically determine the correctness of signature $\sigma_{B\leftarrow A}$. The smart contract is a self-executing piece of code deployed on a blockchain~\cite{buterin2014next}, designed to automatically enforce, verify, or execute terms of an agreement when predefined conditions are met. If the $\sigma_{B\leftarrow A}$ passes the verification, the smart contract will send the pre-stored Crypto with corresponding value to the sender automatically. 

\subsection{Key Features of Our Method}
To summarize, the overall system is demonstrated in Figure~\ref{fig:ecosystem}. The semi-quantum token system combines classical and quantum technologies to create a secure, unclonable, and transferable digital asset. The following features highlight its advantages:
\begin{itemize}
    \item \textbf{Minting:} Quantum tokens are created with inherent uniqueness, making duplication impossible. The bank publishes public information to allow anyone to verify the token's authenticity.
    \item \textbf{Verification:} Tokens can be validated without being destroyed, enabling multiple verifications while preserving the token’s usability.
    \item \textbf{Transfer:} Tokens can be securely transferred between parties, with certified destruction guaranteeing that the original token is invalidated, preventing double-spending or forgery. Such transfer supports multi-channel, including offline and online, centralized or decentralized ways.
\end{itemize}

\section{Experiment}

To validate our proposed method, we conducted an experiment using a real quantum computer and blockchain network. Given the current limitations of noisy intermediate-scale quantum (NISQ) devices~\cite{brooks2019beyond} and the absence of scalable, error-free quantum computers with quantum RAMs, we adapted the design of quantum tokens to fit the capabilities of modern quantum hardware while retaining their core properties. Specifically, we used QuaFu, a quantum cloud platform, to generate quantum token units and perform quantum signatures. In parallel, we deployed a smart contract on Sepolia, the Ethereum testnet~\cite{sepolia}, to verify the signatures and enable secure token transfers automatically.

\subsection{Minting and Signing Quantum Tokens}

We first mint quantum tokens. Each token was simplified to use $\lambda=4$ qubits (a parameter indicating the token's size) to remain compatible with current quantum hardware. We omitted the homomorphic encryption component in this demonstration due to limited qubit resources, but this does not affect the validity of the protocol for the purposes of the experiment. 

A quantum token unit is represented as a $4$-qubit quantum state, defined as:
\begin{equation}
    \ket{\$} := (\ket{0000} + \ket{S_0} + \ket{S_1} + \ket{S_0 \oplus S_1}),
\end{equation}
where $S_0$ and $S_1$ are basis states of the system. The minting process generates these states and appends their corresponding quantum signatures:
\begin{equation}
    S_0 := \sigma_{\ket{\$}, 0} \leftarrow Sign(\ket{\$}, 0), \quad S_1 := \sigma_{\ket{\$}, 1} \leftarrow Sign(\ket{\$}, 1).
\end{equation}

\begin{figure}[h!]
    \centering
    \includegraphics[width=1\linewidth]{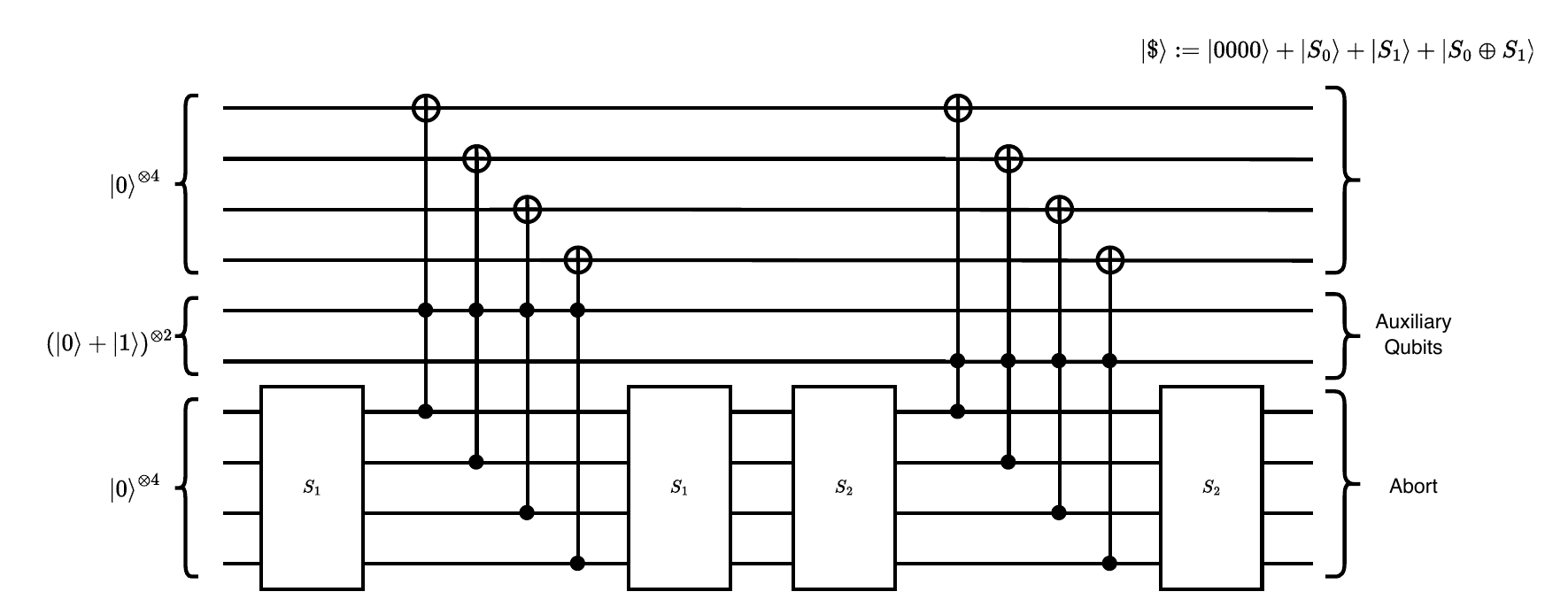}
    \caption{Quantum Circuit for Minting and Signing Tokens}
    \label{fig:mint_demo}
\end{figure}

Figure~\ref{fig:mint_demo} shows the quantum circuit used to mint and sign the quantum token. The circuit is structured as follows:
\begin{itemize}
    \item The first register contains 4 qubits, which form the quantum token unit $\ket{\$}$.
    \item The second register contains 2 auxiliary qubits, which are not involved in subsequent processes.
    \item The third register contains 4 qubits, which are discarded after the minting process.
\end{itemize}

This setup ensures that the quantum token is securely minted and signed, while adhering to the constraints of current quantum hardware.

\subsection{Transferring Tokens via Blockchain Smart Contracts}

To simulate the transfer of quantum tokens, we deployed smart contracts on the Sepolia Ethereum testnet\footnote{The main contract address is: \texttt{0x37F341aeA9Aa504a2b324EB4752189ba04B5244C}. Use method \texttt{getUpdateInfo} for details of the whole information.}. The demonstration involved three key participants:
\begin{itemize}
    \item \textbf{Bank}: The Bank acts as the oracle and deploys a smart contract called \texttt{QuCoin}, which stores information about token ownership and value. Only the Bank can write to this database, while other participants (e.g., Sender, Receiver) and smart contracts can read from it. The \texttt{QuCoin} smart contract provides:
    \begin{itemize}
        \item \texttt{getOracle(tokenID)}: Retrieves the oracle information for a specific quantum token.
        \item \texttt{getValue(tokenID)}: Retrieves the value of a specific quantum token.
    \end{itemize}
    
    \item \textbf{Receiver (Rec)}: The Receiver deploys a smart contract called \texttt{Transfer} to handle incoming transactions. This contract stores the index of a dummy token (\texttt{id\_B}) and an amount of Sepolia cryptocurrency (denoted as $x$), which is deposited and cannot be withdrawn.

    \item \textbf{Sender (Sen)}: The Sender signs the quantum token $A$ (with value $x_A$) using the dummy token index (\texttt{id\_B}) to generate a signature $\sigma_{B \leftarrow A}$. The Sender then calls the \texttt{sign(signerID, signature)} function on the \texttt{Transfer} contract to initiate the transfer.
\end{itemize}

The smart contract automates the transfer by verifying:
\begin{itemize}
    \item The value of token $A$ ($x_A$) is greater than or equal to the deposited cryptocurrency value ($x$).
    \item The signature $\sigma_{B \leftarrow A}$ is valid.
\end{itemize}

If both conditions are met, the smart contract executes the following actions:
\begin{itemize}
    \item Transfers the deposited cryptocurrency to the Sender.
    \item Sends the signature $\sigma_{B \leftarrow A}$ to the Receiver.
\end{itemize}

This process demonstrates how quantum tokens can be securely transferred and verified using blockchain technology, ensuring that all protocol rules are enforced without relying on centralized intermediaries. By integrating quantum and blockchain technologies, this demonstration showcases the feasibility of a decentralized, secure, and scalable multi-channel currency ecosystem.

\section{Conclusion}

Our research offers significant theoretical implications for future information systems studies by advancing the discourse on novel IT artifacts, particularly quantum tokens. Building upon foundational work in quantum money~\cite{wiesner1983conjugate, aaronson2009quantum} and recent advancements in quantum tokens~\cite{shmueli2022public, shmueli2022semi}, we address critical limitations regarding quantum resource and channel requirements. Unlike prior approaches that assume full quantum computational capabilities for all users~\cite{shmueli2022semi}, our work demonstrates how semi-quantum systems can reduce these demands while still ensuring essential properties like unclonability, certified destruction, and decentralized verification. This shift broadens the accessibility of quantum-enabled systems, making them more practical for real-world adoption.

Additionally, this study contributes to the theoretical understanding of digital tokens, including their use in central bank digital currencies~\cite{agur2022designing}, tokenized deposits, and stablecoins~\cite{garratt2023stablecoins}, by illustrating how quantum technology can enhance attributes like security, scalability, and privacy~\cite{jin2022cev, jin2024scalability}. By minimizing the quantum resources required for network nodes and channels~\cite{wiesner1983conjugate, shmueli2022public}, our research establishes a framework for hybrid quantum-classical systems that balances advanced quantum capabilities with practical implementation constraints. Theoretically, this work invites future research to explore new use cases, empirical implementations, and the integration of quantum mechanics with traditional IT infrastructures, setting the stage for a transformative shift in secure and efficient information systems.

We provide significant practical implications by addressing the core limitations of offline payment systems through our multi-channel semi-quantum currency protocol. Unlike traditional methods, which are prone to counterfeiting, tampering, and reliance on centralized trust, our protocol leverages quantum technology to ensure unclonability, certified destruction, and decentralized verification. Its multi-channel design—supporting face-to-face, no-middleman remote, and on-chain transfers—provides flexibility for diverse use cases while preventing double-spending and enhancing transaction security. By enabling reliable offline payments without requiring continuous connectivity or specialized hardware, this protocol fosters financial inclusion in underserved regions and establishes a scalable, secure, and transformative framework for offline financial systems.

\appendix

\section{Cryptography Tools}\label{crptography tools}

\subsection{Leveled Hybrid Quantum Fully Homomorphic Encryption (QFHE)}

Leveled QFHE is a critical cryptographic tool in the semi-quantum token protocol, based on the learning-with-errors (LWE) problem \cite{regev2009lattices}, a widely recognized quantum-resistant assumption \cite{peikert2014lattice}. It builds on classical Fully Homomorphic Encryption (FHE) \cite{gentry2009fully, gentry2013homomorphic} and extends its functionality to quantum data, enabling secure outsourcing of quantum computations to a semi-trusted third party without exposing plaintext quantum information.

Classical FHE allows secure computation on encrypted data through the following steps:
\begin{enumerate}
    \item A key pair, public key ($pk$) and secret key ($sk$), is generated by the delegator.
    \item A private message $m$ is encrypted into ciphertext $Enc_{pk}(m)$ using $pk$.
    \item A Boolean function $f$ is evaluated on the encrypted data using $Eval(\cdot, f)$, producing $Enc_{pk}(f(m))$, all without decrypting the data.
    \item The delegator decrypts the result with $sk$ to retrieve $f(m)$.
\end{enumerate}

Quantum FHE (QFHE) extends these principles to secure quantum data, protecting both classical and quantum information. Its process is as follows:
\begin{enumerate}
    \item The delegator encrypts a quantum state $\ket{\psi}$ using a Quantum One-Time Pad (QOTP), resulting in $\ket{\psi}^{(x, z)}$, where $(x, z)$ are random bitstrings.
    \item The bitstrings $(x, z)$ are further encrypted using a quantum-capable homomorphic encryption scheme, producing $Enc_{pk}(x, z)$.
    \item The delegatee homomorphically evaluates a quantum circuit $C$ on the encrypted state, producing:
    \begin{itemize}
        \item An encrypted quantum state $(C(\ket{\psi}))^{(x', z')}$ under a new QOTP.
        \item A classical ciphertext $ct_{x', z'}$, encoding the updated QOTP keys $(x', z')$.
    \end{itemize}
    \item The delegator decrypts $ct_{x', z'}$ using $sk$, recovers $(x', z')$, and decrypts the final quantum state to retrieve $C(\ket{\psi})$.
\end{enumerate}

QFHE ensures privacy and security for quantum computations, leveraging the robustness of LWE to resist quantum adversaries.

\begin{figure}[h!]
    \centering
    \begin{subfigure}{\textwidth}
        \includegraphics[width=1.0\textwidth]{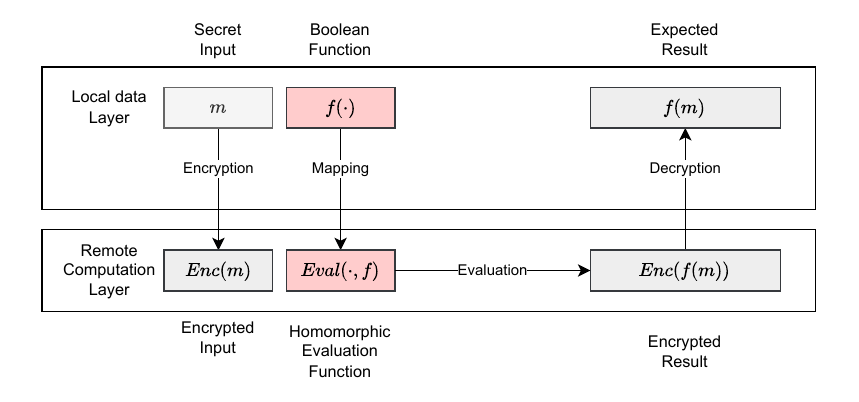}
        \caption{Fully Homomorphic Encryption (FHE)}
        \label{fig_FHE}
    \end{subfigure}
    \begin{subfigure}{\textwidth}
        \includegraphics[width=1.0\textwidth]{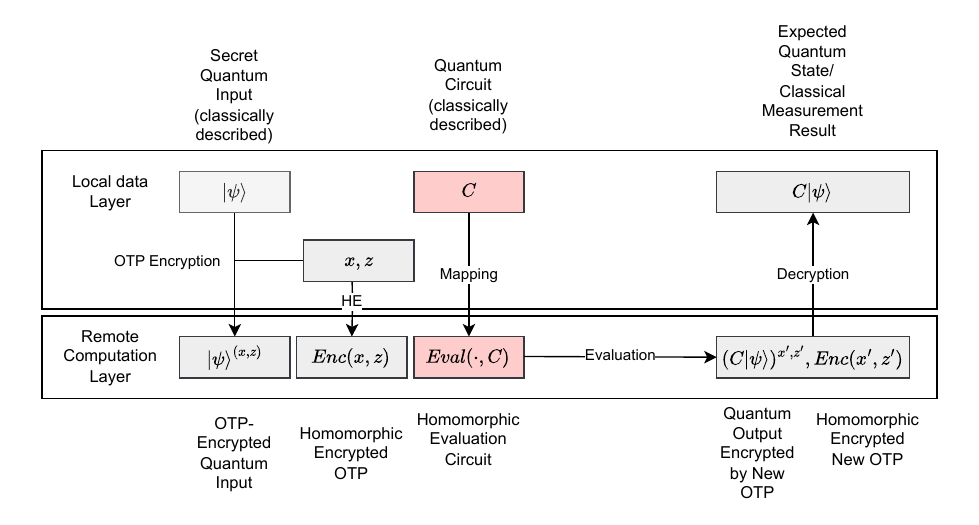}
        \caption{Fully Quantum Homomorphic Encryption (QFHE)}
        \label{fig_QFHE}
    \end{subfigure}
    \caption{Dataflow in Homomorphic Encryption}
    \label{fig_FHE_QFHE}
\end{figure}

\subsection{Quantum Lighting}

A key property of QFHE is its non-deterministic behavior for circuits containing Toffoli gates, a fundamental quantum logic gate. Specifically, the transformation $(x, z) \rightarrow (x', z')$ becomes probabilistic. This unpredictability is essential for cryptographic applications requiring unique, non-reproducible outputs.

We claim that for a quantum circuit $Q^*$ and a QFHE encryption pair $(\ket{\psi}^{(x, z)}, ct_{(x, z)})$, applying the QFHE evaluation function twice almost always yields different outputs. Formally, for two executions of $QHE.Eval$:
\[
(Q^* \ket{\psi})^{(x_A, z_A)}, ct_{(x_A, z_A)} \quad \text{and} \quad (Q^* \ket{\psi})^{(x_B, z_B)}, ct_{(x_B, z_B)},
\]
the probability of obtaining identical outputs is negligible. This ensures the uniqueness of quantum outputs, a critical feature for quantum token generation.

\section{Semi-Quantum Token Processes}\label{semi-quantum_token}

\subsection{Mint}

A semi-quantum token comprises classical information and quantum token units. The classical part includes $(pk_1, pk_2, ..., pk_\lambda, x, id)$, where each $pk_n$ is the public key of the $n$-th token unit, $x$ is the value assigned by the $Bank$, and $id$ is a unique $\lambda$-length bit string. The quantum part is a set of quantum token units: $(\ket{\$}_1, \ket{\$}_2, ..., \ket{\$}_\lambda)$. 

The minting process allows a receiver to obtain a quantum token unit $\ket{\$}$ securely, while the bank retains verification data. Algorithm~\ref{alg: mint} outlines the minting procedure.

\begin{algorithm}[h!]
\footnotesize
\caption{Mint Quantum Token Unit}\label{alg: mint}
\begin{algorithmic}[1]
\REQUIRE A classical bank $Bank$, a quantum receiver $Rec$, and a security parameter $\lambda$
\ENSURE $Rec$ outputs quantum token unit $\ket{\$}$ and ciphertext $ct_{x,z}$, $Bank$ retains a verification oracle $(O_{S_0+x}, O_{S_0+x+w}, O_{S^{\perp}+z})$ and its database address $pk$
\STATE $Bank$ generates a random $\frac{\lambda}{2}$-dimensional subspace matrix $M_S \in \{0,1\}^{\frac{\lambda}{2}\times \lambda}$
\STATE $Bank$ encrypts $M_S$ with a random OTP key $p_x$ and encrypts $p_x$ using a quantum homomorphic encryption (QHE) key, producing $M_S^{(p_x)}$ and $ct_{p_x}$
\STATE $Bank$ sends $(M_S^{(p_x)}, ct_{p_x})$ to $Rec$
\STATE $Rec$ evaluates the row-span circuit homomorphically to generate $(\ket{\psi}^{(x,z)}, ct_{x,z})$
\STATE $Rec$ returns $ct_{x,z}$ to $Bank$, which decrypts $(x, z)$
\IF {$x \notin S$}
    $Bank$ publishes obfuscated oracles $(O_{S_0+x}, O_{S_0+x+w}, O_{S^{\perp}+z})$ at a unique address $pk$
\ELSE
    $Bank$ restarts the protocol
\ENDIF
\end{algorithmic}
\end{algorithm}

\subsection{Verification}

A quantum token unit $\ket{\$}$ can be verified non-destructively, meaning the quantum state remains intact if verification succeeds. The public verification oracle $(O_{S_0+x}, O_{S_0+x+w}, O_{S^{\perp}+z})$ ensures that only valid token units pass verification. Algorithm~\ref{alg:Quantum Verification} describes the process.

\begin{algorithm}[h!]
\footnotesize
\caption{Quantum Token Unit Verification}\label{alg:Quantum Verification}
\begin{algorithmic}[1]
\REQUIRE $Rec$ holds $\ket{\$}$ and public oracle $(O_{S_0+x}, O_{S_0+x+w}, O_{S^{\perp}+z})$
\ENSURE $Rec$ outputs a bit $m$ indicating the verification result
\STATE $Rec$ applies oracle $O$ to test $(O_{S_0+x} \vee O_{S_0+x+w})$ and $(O_{S^{\perp}+z})$, measuring the result $m$
\IF {$m=0$}
    Verification fails
\ELSE
    $Rec$ repeats the test for all oracles
\IF {All tests pass}
    $Rec$ accepts the token
\ELSE
    $Rec$ rejects the token
\ENDIF
\ENDIF
\end{algorithmic}
\end{algorithm}

\subsection{Transfer}\label{sec:transfer}

Quantum token units can be destroyed irreversibly during transfer. Each unit collapses into one of two non-overlapping subspaces, $S_0$ or $S_1$, based on a classical control bit $b$. The measurement result $\sigma_{\ket{\$},b} \leftarrow Sign(\ket{\$},b)$ serves as the token's signature, proving:
1. The token collapsed into subspace $S_b$
2. The token was measured and cannot return to its quantum state

Each quantum token unit can only be signed once. To transfer token $A$ (value $x_A$) to a receiver’s dummy token $B$ (value $x_B=0$), the sender signs each unit of $A$ using $B$'s unique identifier $id_B$. The resulting signature $\sigma_{B \leftarrow A}$ confirms that $x_A$ has been transferred to $B$. Algorithm~\ref{alg:Quantum Signing} details the signing process.

\begin{algorithm}[h!]
\footnotesize
\caption{Quantum Signing}\label{alg:Quantum Signing}
\begin{algorithmic}[1]
\REQUIRE A quantum sender $Sen$ holds $\ket{\$}$ and a classical bit $b$
\ENSURE $Sen$ outputs a signature $\sigma_{\ket{\$},b}$, and $\ket{\$}$ is destroyed
\STATE $Sen$ applies oracle $(O_{S_0+x+bw})$ to $\ket{\$}$ and measures the result $m$
\IF {$m=1$}
    $Sen$ measures $\ket{\$}$ to obtain $\sigma_{\ket{\$},b}$
\ELSE
    $Sen$ verifies with $(O_{S^{\perp}+z})$ and repeats the process
\ENDIF
\end{algorithmic}
\end{algorithm}

\bibliography{refs}

\end{document}